# It’s Not the Tool, It’s the Task:
## A Framework for Cognitively Activated AI Augmentation in Physics Instruction

Jochen Kuhn[1], Stefan Küchemann[1], David J. Rakestraw[2], & Patrik Vogt[3]

[1]Ludwig-Maximilians-Universität München (LMU Munich), Faculty of Physics/Chair of Physics Education Research, 80539 Munich, Germany

[2]Lawrence Livermore National Laboratory, Livermore, CA 94550

[3]Heidelberg University of Education, Institute of Physics, Heidelberg, Germany

### 1. Introduction: The Boiling Frog Problem in Physics Education

Multiple generative AI foundation models can currently solve introductory physics problems at a level comparable to expert human performance. Kortemeyer (2026) reports that GPT-5 Thinking attains scores of 27/30 and 25/30 on comprehensive physics finals, scores 92.3% on the TUG-Kv4.0 representation translation test, and categorizes physics problems by solution method rather than surface features (a pattern previously associated with expert cognition).[1] Performance on standardized examinations in adjacent high-stakes domains (legal bar examinations, medical licensing) similarly exceeds average human performance. The models continue to improve at astounding rates, and as of summer 2026 (GPT-5.6, Claude Fable, Gemini 3.6 Flash), most standard introductory kinematics, thermodynamics, and electromagnetism problems can be solved almost instantaneously.

Small accommodations, allowing AI-generated data visualization, supporting AI-assisted explanations with complementary perspectives, or shifting emphasis toward assessing understanding without relying on final answers, can accumulate over time. The danger lies not in incorrect results, but in the quiet erosion of students’ engagement in epistemic practices (the ways of thinking, predicting, and evaluating evidence that define scientific reasoning)[1]. Kortemeyer describes this as the "boiling frog problem" of physics education[1]: just as the frog fails to recognize gradually increasing danger, educational systems may adapt so incrementally to AI capabilities that their foundational epistemic goals are no longer questioned, until it is too late to reverse the adaptation. His analysis poses the diagnostic question: *when is it time to jump?*

The present paper proposes a constructive answer to that question at the level of individual classroom design. The central challenge of AI in physics education is instructional design, not cheating or tool selection. Drawing on research on self-regulated learning, cognitive load, multiple representations, hybrid intelligence, and recent meta-analytic evidence of the value of cognitively activating in context of generative AI[2], we propose AIRIS: a practical framework for cognitively activated learning activities that structures student work before, during, and after AI use. The three-phase structure presented here is a deliberately practice-oriented condensation of a broader, more fine-grained account of how regulation is distributed in hybrid human-AI cognition, developed in a companion theoretical paper; the present article focuses on what teachers can put to work in the classroom. The focus must shift to the principle that **"it’s not the tool, it’s the task."** Although a well-equipped individual will frequently outperform a less-equipped one in many domains, the core issue in education is fundamentally different. The relevant question is not whether AI augments performance (it undeniably does) but whether that augmentation enhances, preserves or displaces the cognitive work essential to learning.

### 2. Learning Physics as Epistemic Practice

Physics education research has long emphasized that learning physics is more than mastering procedures. It involves engaging in practices that include constructing models, coordinating representations, evaluating assumptions, and reflecting on uncertainty.

Several research traditions are particularly relevant for understanding the impact of AI on these practices:

- Self-Regulated Learning (SRL): Students' capacity to plan their approach, monitor their progress, and reflect on outcomes. When AI silently performs key cognitive steps, these regulatory processes may be bypassed rather than supported[4,5].
- Cognitive Theory of Multimedia Learning (CTML): Reducing extraneous cognitive load can support learning, but eliminating *germane* cognitive effort (cognitive effort directed at making sense of the material) undermines understanding[6,7].

- Multiple External Representations (MER): Conceptual understanding emerges from coordinating representations (e.g., graphs, equations, verbal descriptions), not merely translating between them[8-11].
- Hybrid Intelligence: Productive human-AI collaboration requires explicit role differentiation, with epistemic responsibility remaining human rather than delegated to the machine[12,13].

A foundational question underlies this framework: if AI can reliably perform numerical integration, visualization, and error analysis, should student competence in those tasks remain a primary learning objective? We argue that the answer depends on the cognitive level of the operation. Procedural execution, running an integration algorithm or generating a standard plot, may legitimately be delegated to AI, just as calculators replaced manual arithmetic without eliminating the need to understand what a calculation means. But the capacity to *predict* what a result should look like before AI generates it, to *evaluate* whether the AI output is physically meaningful, and to *diagnose* discrepancies between expectation and output: these remain irreducibly human epistemic tasks. Competence in coordinating representations across modalities, reading what a *v-t*-diagram implies for an *s-t*-diagram, judging whether a fitted model is physically reasonable, cannot be delegated without undermining the learning goals that the course exists to serve. AIRIS is designed precisely to preserve these higher-order operations while allowing procedural delegation.
These perspectives converge on a key insight: AI use must be deliberately structured around student activity in a cognitively activated way if learning goals are to be preserved.

## 3. AIRIS: Activate – Inquire – Reflect – with Intelligent Support

We propose a simple but powerful three-phase instructional framework for AI-augmented science education that structures student activities before, during, and after AI use considering the educational practices mentioned above. The framework is not tied to a specific tool. It focuses on *when* and *how* AI enters the learning process in three phases: Activation, Inquiry, and Reflection with Intelligent Support (AIRIS). Each phase bundles several finer-grained cognitive operations, for instance, anticipating an outcome, interrogating an AI output, and integrating representations, that a more detailed theoretical treatment separates out; for classroom use we group them into the three phases that teachers can plan and observe directly.

### 3.1 Student Activities *Before* AI Use: Epistemic Grounding – Activate

Before AI is introduced into the workflow, students engage in activities that activate prior knowledge and establish expectations. Typical tasks include:

- exploring experimental design options,
- sketching expected relationships between measured quantities (graphs, field patterns, or system diagrams),
- making qualitative predictions about physical behavior (e.g., proportional, inverse, exponential, or periodic relationships),
- identifying relevant variables, parameters, and constraints,
- estimating magnitudes, ratios, or characteristic scales.

These activities serve two purposes. First, they encourage students to construct initial mental models. Second, they create a reference point against which AI-generated representations can later be evaluated. From an SRL perspective, this phase supports planning; from a CTML perspective, it prepares students for meaningful cognitive processing.

### 3.2 Student Activities *During* AI Use: Guided Delegation – Inquire

During AI use, computational or representational tasks may be delegated, but epistemic judgment remains with students. In introductory kinematics labs, for example, students may use AI-supported tools to:

- generate visualizations or mathematical models from experimental data,
- visualize patterns across multiple trials or conditions,
- perform routine calculations.

Crucially, students are not asked to accept AI outputs at face value. Instead, they compare AI-generated results with their own predictions, annotate discrepancies, and question assumptions such as fitting ranges or smoothing choices. In this phase, AI functions as a *computational and representational partner*, consistent with a hybrid intelligence perspective.

3.3 Student Activities *After* AI Use: Epistemic Responsibility – Reflect

Learning is consolidated after AI use through interpretation and reflection. Students are asked to:

- explain the physical meaning of model parameters, coefficients, or derived quantities,
- compare results with theoretical expectations or literature values,
- diagnose sources of uncertainty, systematic error, or model limitations,
- reflect explicitly on what the AI contributed and what remained their responsibility.

This phase directly addresses the risk of "hollowed-out" understanding by making epistemic responsibility explicit. From an SRL standpoint, it emphasizes monitoring and reflection; from an MER perspective, it requires students to coordinate representations meaningfully.

3.4 Role of the teacher

Each phase of AIRIS places specific demands on the teacher, not as the central actor but as the designer and guardian of the task structure. In Phase 1, the teacher selects or constructs tasks that require genuine prior-knowledge activation rather than surface recall. In Phase 2, the teacher monitors whether students are critically comparing AI outputs against their own predictions or simply accepting them. In Phase 3, the teacher orchestrates reflection so that epistemic responsibility is explicitly located with the learner. These are forms of pedagogical judgment that AI cannot supply. What AIRIS asks of the teacher, in short, is to protect the integrity of the task at every stage.

AIRIS shares the three-phase structure of established inquiry cycles such as 5E[14] and Predict-Observe-Explain[15]. Its specific contribution is the integration of AI as an explicit representational and computational partner within the Inquire phase, with deliberate design principles for what must remain the student's cognitive responsibility, and the addition of a Reflect phase that requires students to explicitly account for what the AI contributed and what they themselves supplied. AIRIS is therefore an AI-specific adaptation of existing inquiry cycles, not a replacement for them.

## 4. Classroom Example: Elevator Ride in the Shard – Analyzing Real Motion Using Acceleration Data and AI

In the following, we would like to illustrate the didactic framework using a concrete example (further examples can be found in[16-18]). AIRIS is designed for inquiry-based instruction and problem-solving settings, but it does of course not replace direct instruction. The conceptual knowledge required to make meaningful predictions in Phase 1 (understanding of acceleration, numerical integration, the relationship between kinematic quantities) must be established through prior instruction. AIRIS structures what happens when students explore, apply and extend that knowledge in AI-augmented settings.

During a downward elevator ride in the Shard (London) from floor 34 to the ground floor, a smartphone was used to record the acceleration as a function of time. Throughout the entire ride, the smartphone was held in a vertical orientation against the elevator wall (Fig. 1). Therefore, only the $y$-component of the acceleration sensor is meaningful, as it represents the vertical motion of the elevator. No access to the Shard or any moving vehicle is required: the dataset is provided with this paper, and the same AIRIS structure applies to any phenomenon that can be measured, modelled, and interpreted, e.g., a smartphone recording a bouncing ball, a rolling cylinder, or a pendulum in a school courtyard.

The goal of this task is to characterize the elevator's motion based on real acceleration data[19,20]. Artificial intelligence is used in a targeted and limited way. The task explicitly follows the three-phase framework for AI-augmented physics learning[17]. The three phases are presented in a format that resembles a student guide to the investigation. This provides a concrete illustration of how to implement AIRIS in the classroom.

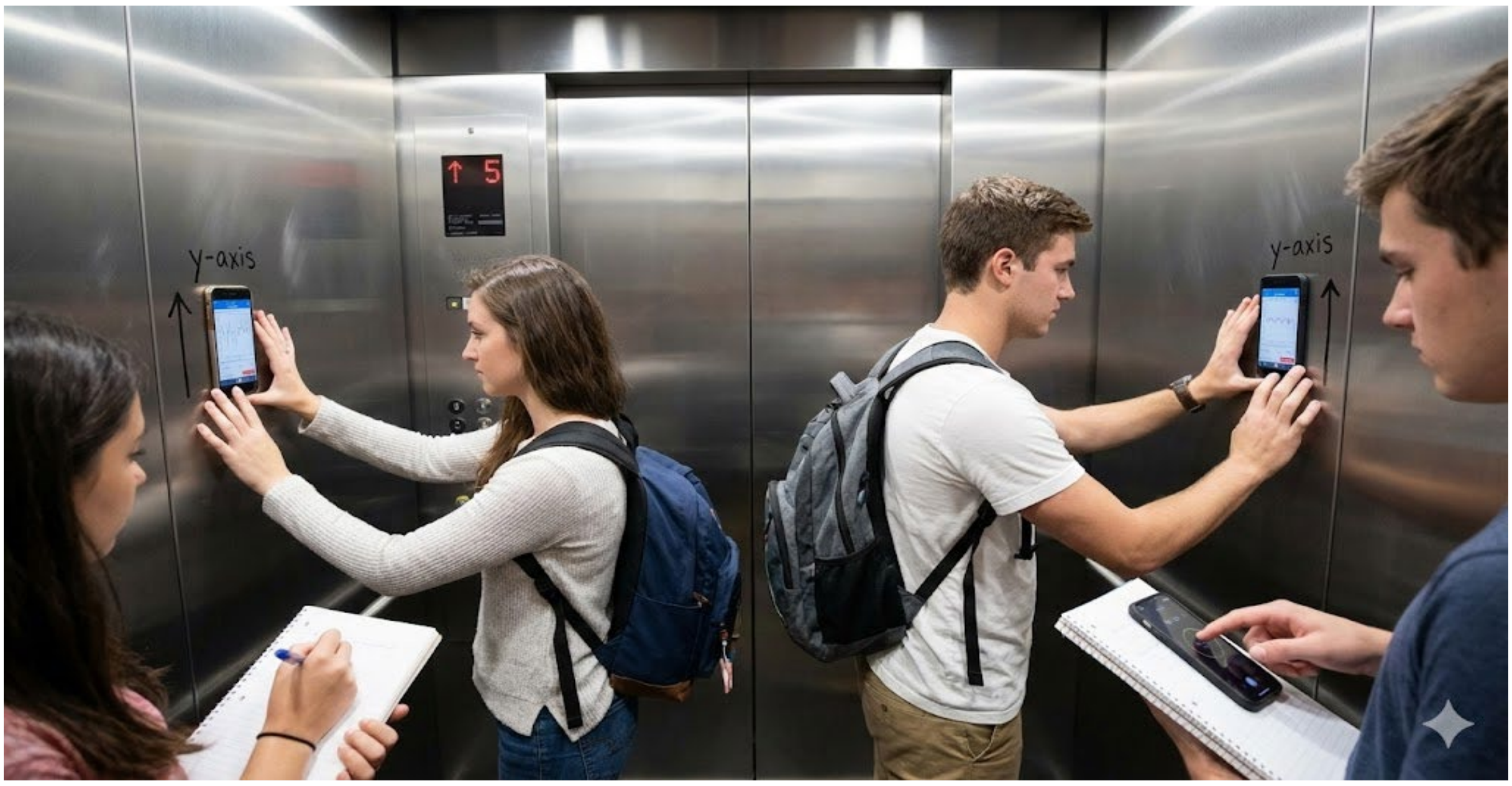

Fig. 1: Illustration of the experimental setup generated by Gemini: Students experimenting in the elevator.

## 4.1 Phase 1: Before AI Use – Activate

Before using AI, you will predict and qualitatively describe the expected motion of the elevator. The goal of this phase is to develop your own ideas and record expectations that will later serve as a reference for comparison.

**A.** Qualitative Modeling
   a. For a typical downward elevator ride, sketch the following:
      i. an *a*-*t* diagram (acceleration vs. time),
      ii. a *v*-*t* diagram (velocity vs. time),
      iii. an *s*-*t* diagram (position vs. time).
   b. Divide the motion into meaningful phases and mark these phases in all three diagrams.

**B.** Expectations Regarding the Measurement
   a. What characteristic features do you expect in the *a*-*t* diagram of a real elevator ride?
      i. sign of the acceleration,
      ii. typical order of magnitude,
      iii. duration of individual phases.
   b. Justify your expectations using physical reasoning.

**C.** Relationship Between Representations
   a. Explain qualitatively how a *v*–*t* diagram and an *s*–*t* diagram can be obtained from an *a*-*t* diagram.
   b. When moving from *a*-*t* to *v*-*t* and *s*-*t* diagrams, not only the representation changes, but also the information content. Describe:
      i. what can be seen more clearly in the later diagrams,
      ii. what is seen less clearly or no longer at all,
      iii. why these diagrams depend more strongly on assumptions.

These sketches and considerations will later serve as a reference for evaluating the AI outputs.

## 4.2 Phase 2: During AI Use – Inquire

During AI use, computational and representational steps are delegated to the AI. Your task is to compare the generated results with your expectations and to critically examine the assumptions involved.

**A.** Visualization of the Measurement Data: Upload the acceleration dataset and ask the AI to generate an *a*-*t* diagram of the *y*-component, marking notable regions or phases in the diagram (Figure 2, top panel). Then estimate the maximum speed of the elevator without using AI.

**B.** Deriving Additional Quantities**:** Ask the AI to determine a *v*-*t* curve from the acceleration data and to calculate an *s*-*t* curve from the *v*-*t* data (Figure 2, middle and lower panels). Does the maximum speed obtained match your estimate from the previous task?

**C.** Critical Comparison: Compare the diagrams generated by the AI with your sketches from Phase 1. Where do they agree, and where do they differ?

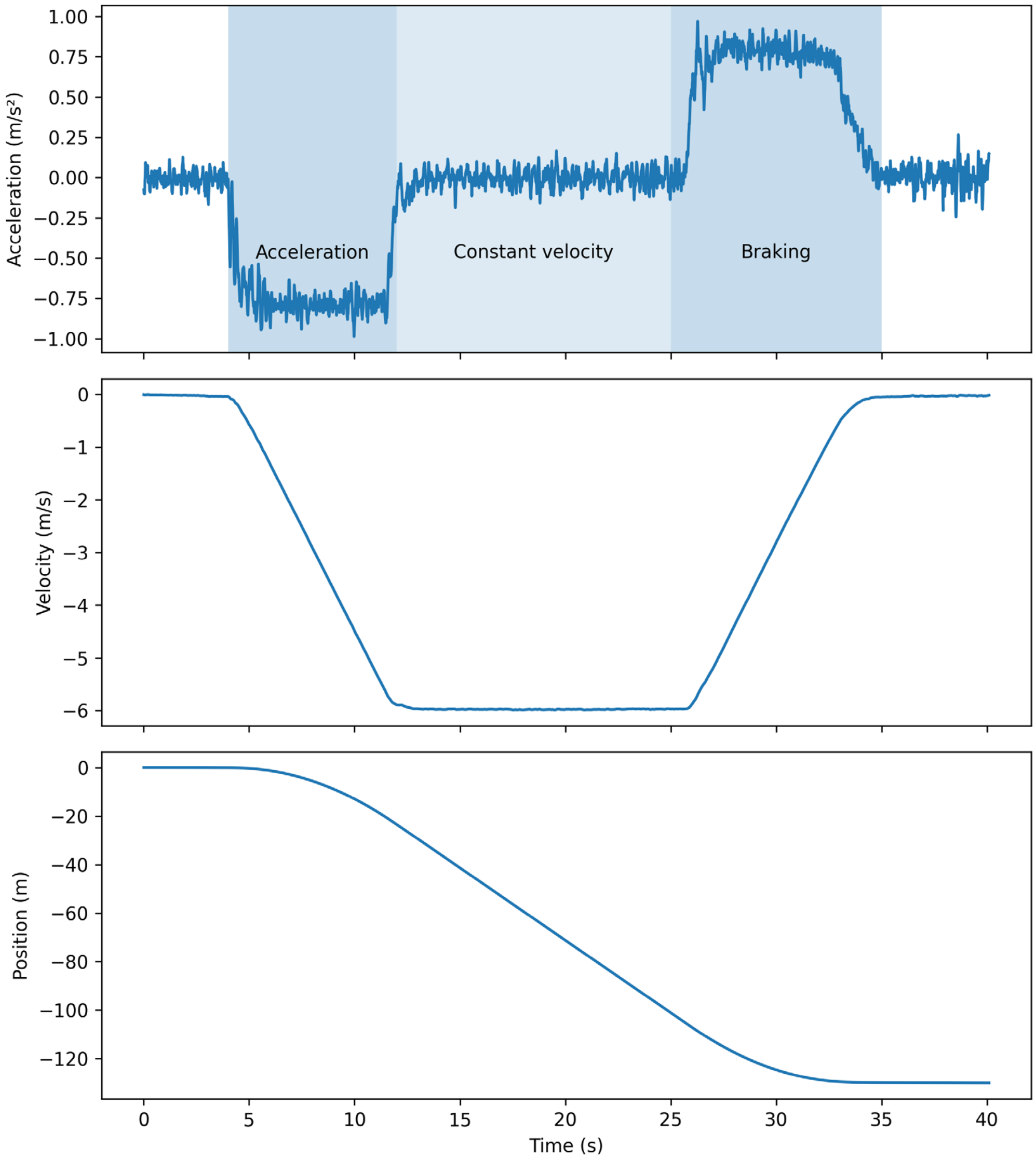

Fig. 2: Vertical acceleration $a(t)$ of the elevator during a downward ride in the Shard (London). The data were corrected for a systematic sensor offset, which was determined as the mean acceleration during the constant-velocity interval from 15 s to 20 s ($a_{\text{offset}}$ = 0.042 m/s$^2$). The colored background indicates the three motion phases—acceleration, constant velocity, and braking. From the corrected acceleration data (top), the corresponding velocity $v(t)$ (middle) and position $s(t)$ (bottom) were obtained by numerical integration and are shown together for comparison.

## 4.3 Phase 3: After AI Use – Reflect

After using AI, the generated representations are interpreted physically and related to one another. The goal of this phase is to place the results in context and to assess their physical explanatory power.

**A.** Physical Interpretation

a. Explain the meaning of the slopes in the *v*-*t* diagram and in the *s*-*t* diagram.

b. Determine the total distance traveled by the elevator from the *s*-*t* diagram and briefly explain how this value is obtained from the representation.

**B.** Consistency Check of the Representations

c. Are the *a-t*, *v-t*, and *s-t* diagrams physically consistent with one another?
d. Check the plausibility of the distance determined from the *s–t* diagram. What average floor height follows from this result?
e. If necessary, identify existing errors, explain them physically, and correct them by re-prompting the AI.

**C.** Model Limitations

e. Discuss typical problems that arise when integrating real acceleration data (e.g., sensor offset, noise, drift).
f. Which parts of the elevator ride can be modelled particularly reliably, and which parts less so?

**D.** Written Reflection (to be submitted before the end of the inquiry session)

Write a brief reflection of 150–200 words addressed to your teacher. Your reflection should answer:

g. What did the AI contribute to this investigation, and what did you contribute?
h. What can you now explain, draw, or calculate about elevator motion that you could not before this task?
i. Where are you still uncertain, and what would you need to check or measure to resolve that uncertainty?

This written reflection serves the function of a traditional inquiry report: it requires you to articulate your reasoning, justify your interpretations, and take responsibility for your conclusions - while additionally making the human-AI division of cognitive work explicit.

## 5. Ethics by Design

The presented AIRIS framework offers a practical classroom-level response to the boiling frog problem identified by Kortemeyer[1]. Where Kortemeyer's analysis operates at the level of the physics education system, documenting the gradual erosion of epistemic standards as AI capabilities advance, AIRIS addresses the analogous risk at the level of individual instructional design. Each accommodation appears reasonable in isolation and it is in line with recent meta-analytic findings from Boolzen and colleagues[2]. Allowing AI-generated plots here, AI-assisted explanations there. The cumulative effect, however, can be a hollowing-out of epistemic engagement that no single decision made visible. AIRIS prevents this by making prediction, evaluation, and reflection non-delegable components of every task.

The ethical demands this places on teachers are distinct from the instructional design question. Teaching physics entails the responsibility to foster students' epistemic competencies: the capacity to model, evaluate, and interpret physical phenomena independently. The German Ethics Council warns that AI should not be used in cases where it leads to a degeneration of relevant competencies of learners[21]. Recent empirical work documents the risks of reduced cognitive activity and critical reflection on one's own reasoning as well as uncritical acceptance of AI outputs[3,22–26]. AIRIS addresses these risks through design, not prohibition.

Implementing AIRIS effectively presupposes that teachers can themselves evaluate AI outputs in their domain, identify plausible errors, and design tasks that resist AI completion. Current teacher preparation programs have not yet systematically addressed these competencies. Teachers should therefore communicate transparently with their students about the epistemic demands and limitations of AI use, and with colleagues and administrators about the instructional strategies they are adopting, so that AI integration is collectively deliberate rather than individually improvised.

## 6. Conclusion

AI is already impacting introductory physics education and will play an increasingly important role as AI tools improve. The critical question is whether it will quietly replace students' epistemic work or be used to make that work more visible and deliberate. The AIRIS framework structures student activities before, during, and after AI use, enabling physics educators to preserve what matters most: helping students learn to think like scientists in a cognitively activated way and in an AI-rich world.

AIRIS is presented here as a design proposal grounded in established learning science principles, not as an empirically validated intervention. The theoretical account from which these principles are drawn, including the mechanisms by which AI use can quietly erode regulation, and the withdrawal logic invoked below, is developed more fully in ongoing theoretical work. Its three phases (Activate, Inquire, Reflect) are motivated by convergent evidence from research on self-regulated learning, cognitive load, multiple representations, and hybrid intelligence, and illustrated through a concrete classroom example that physics teachers can adapt directly. The elevator example is one instance of a broader class of tasks: any measurement-based investigation in which students supply original observations, predict outcomes before AI engagement, and evaluate AI-generated representations against their own expectations satisfies the AIRIS design logic.

The decisive empirical question is whether students who complete AIRIS-structured tasks retain the capacity to generate, evaluate, and coordinate representations when AI support is removed. Answering it requires experimental designs that go beyond classroom observation. This is the withdrawal condition: what students can still explain and produce once the model is no longer available. It requires comparing AI-supported and AI-withdrawn performance at sufficient scale to distinguish durable competence from scaffolded performance. Addressing this question is an open problem for the science of learning, and one that determines whether instructional frameworks like AIRIS succeed at their central goal.

**Acknowledgements**

We thank Peter Sander for recording the measurement data and for providing the dataset.

**References**

1. Kortemeyer, G. (2026). The boiling-frog problem of physics education. Phys. Teach. 64, 8–12.
2. Boolzen, C., Kuhn, J., Flegr, S., Rott, E.-M., Stausberg, N., & Küchemann, S. (2026) Evidence of Impact and Interpretational Limits of Generative AI in STEM education-A Systematic Review and Meta-Analysis on Cognitive Learning Outcomes, *Artif. Intell. Rev.* (accepted)
3. Fan, Y., Tang, L., Le, H., Shen, K., Tan, S., Zhao, Y., ... & Gašević, D. (2025). Beware of metacognitive laziness: Effects of generative artificial intelligence on learning motivation, processes, and performance. British Journal of Educational Technology, 56(2), 489-530.
4. Zimmerman, B. J. (2002). Becoming a self-regulated learner. *Theory Into Practice, 41*(2), 64–70.
5. de Mooij, S., Lämsä, J., Lim, L., Aksela, O., Athavale, S., Bistolfi, I., ... & Molenaar, I. (2025). A Systematic Review of Self-Regulated Learning through Integration of Multimodal Data and Artificial Intelligence. Educational Psychology Review, 37(2), 54.
6. Sweller, J., Ayres, P., & Kalyuga, S. (2019). *Cognitive Load Theory*. Springer.
7. Mayer, R. E. (2020). *Multimedia Learning* (3rd ed.). Cambridge University Press.
8. Ainsworth, S. (2006). DeFT: A conceptual framework for considering learning with multiple representations. *Learning and Instruction, 16*(3), 183–198.
9. Becker, S., Klein, P., Gößling, A., & Kuhn, J. (2020). Investigating dynamic visualizations of multiple representations using mobile video analysis in physics lessons: effects on emotion, cognitive load and conceptual understanding. Zeitschrift für Didaktik der Naturwissenschaften, 26(1), 123-142.
10. Klein, P., Müller, A., & Kuhn, J. (2017). Assessment of representational competence in kinematics. Physical Review Physics Education Research, 13(1), 010132.
11. Opfermann, M., Schmeck, A., Fischer, H.E. (2021). Multiple Representations and Learning Physics. In: Fischer, H.E., Girwidz, R. (eds) Physics Education. Challenges in Physics Education (pp. 175–199). Springer, Cham
12. Cukurova, M. (2025). The interplay of learning, analytics and artificial intelligence in education: A vision for hybrid intelligence. British Journal of Educational Technology, 56(2), 469-488.

13. Molenaar, I. (2022). Towards hybrid human-AI learning technologies. European Journal of Education, 57(4), 632-645.
14. Bybee, R. W. (1997) Achieving scientific literacy: From purposes to practices. Portsmouth, NH: Heinemann.
15. White, R. T., & Gunstone, R. F. (1992). Probing Understanding. Great Britain: Falmer Press.
16. Küchemann, S., Vogt, P., Rakestraw, D. & Kuhn, J. (2026). AI-based analysis of damped Spring Pendulum. *Phys. Teach. 64*(3), 224-225.
17. Kuhn, J., Rakestraw, D. J., Küchemann, S., & Vogt, P. (2026). AI-supported mini-labs: Combining smartphone-based experiments and multimodal AI. *European Journal of Physics*, *47*(2), 025702.
18. Kim, D., Havarko, T., Küchemann, S., Kuhn, J., Vogt, P. & Rakestraw, D. (2026). Star Trail Analysis with AI Co-Investigator. Phys. Teach. 64, submitted.
19. Kuhn, J., Vogt, P., & Müller, A. (2014). Analyzing elevator oscillation with the smartphone acceleration sensors. The Physics Teacher, 52(1), 55-56.
20. Yi, W. (2025). Exploring elevator operating parameters with smartphone barometers. The Physics Teacher, 63(7), 604-605.
21. Ethikrat, D. (2023). Mensch und Maschine–Herausforderungen durch Künstliche Intelligenz. *Stellungnahme. Berlin*, *14*, 2023.
22. Fernandes, D., Villa, S., Nicholls, S., Haavisto, O., Buschek, D., Schmidt, A., ... & Welsch, R. (2025). AI makes you smarter but none the wiser: The disconnect between performance and metacognition. *Computers in Human Behavior*, 108779.
23. Kosmyna, N., Hauptmann, E., Yuan, Y. T., Situ, J., Liao, X. H., Beresnitzky, A. V., ... & Maes, P. (2025). Your brain on ChatGPT: Accumulation of cognitive debt when using an AI assistant for essay writing task. *arXiv preprint arXiv:2506.08872*, *4*.
24. Krupp, L., Steinert, S., Kiefer-Emmanouilidis, M., Avila, K. E., Lukowicz, P., Kuhn, J., ... & Karolus, J. (2024, December). Challenges and opportunities of moderating usage of large language models in education. In Proceedings of the International Conference on Mobile and Ubiquitous Multimedia (pp. 249-254).
25. Krupp, L., Steinert, S., Kiefer-Emmanouilidis, M., Avila, K. E., Lukowicz, P., Kuhn, J., ... & Karolus, J. (2024). Unreflected Acceptance–Investigating the Negative Consequences of ChatGPT-Assisted Problem Solving in Physics Education. In HHAI 2024: Hybrid Human AI Systems for the Social Good (pp. 199-212). IOS Press.
26. Sailer, M., Maier, R., Berger, S., Kastorff, T., & Stegmann, K. (2024). Learning activities in technology-enhanced learning: A systematic review of meta-analyses and second-order meta-analysis in higher education. *Learning and Individual Differences*, *112*, 102446.